\documentclass[twocolumn,showpacs,preprintnumbers,superscriptaddress,pre]{revtex4-1}
\usepackage{amssymb,amsmath,graphicx,epsfig,times}

\begin{document}

\title[Universal geometrical factor of protein conformations as a consequence of energy minimization]%
{Universal geometrical factor of protein conformations as a consequence of energy minimization}%

\date{February 12, 2012}

\author{Ming-Chya Wu}%
\email{mcwu@ncu.edu.tw}%
\affiliation{Research Center for Adaptive Data Analysis, National Central University, Chungli 32001, Taiwan}%
\affiliation{Institute of Physics, Academia Sinica, Nankang, Taipei 11529, Taiwan}%
\affiliation{Department of Physics, National Central University, Chungli 32001, Taiwan}%

\author{Mai Suan Li}%
\affiliation{Institute of Physics, Polish Academy of Sciences, Al. Lotnikow 32/46, 02-668 Warsaw, Poland} %

\author{Wen-Jong Ma}%
\affiliation{Institute of Physics, Academia Sinica, Nankang, Taipei 11529, Taiwan} %
\affiliation{Graduate Institute of Applied Physics, National Chengchi University, Taipei 11605, Taiwan} %

\author{Maksim Kouza}%
\affiliation{Institute of Physics, Polish Academy of Sciences,
Al. Lotnikow 32/46, 02-668 Warsaw, Poland} %
\affiliation{Department of Physics, Michigan Technological University, Houghton, MI 49931, USA}

\author{Chin-Kun Hu}%
\email{huck@phys.sinica.edu.tw}%
\affiliation{Institute of Physics, Academia Sinica, Nankang, Taipei 11529, Taiwan}%
\affiliation{Center for Nonlinear and Complex Systems and Department of Physics,
Chung Yuan Christian University, Chungli 32023, Taiwan}%

\begin{abstract}
The biological activity and functional specificity of proteins depend on their native three-dimensional structures determined by inter-and intra-molecular interactions. In this paper, we investigate the geometrical factor of protein conformation as a consequence of energy minimization in protein folding. Folding simulations of $10$ polypeptides with chain length ranging from $183$ to $548$ residues manifest that the dimensionless ratio ($V/{A \langle r \rangle}$) of the van der Waals volume $V$ to the surface area $A$ and average atomic radius $\langle r \rangle$ of the folded structures, calculated with atomic radii setting used in SMMP [Eisenmenger F., \textit{et. al.}, \textit{Comput. Phys. Commun.}, \textbf{138} (2001) 192], approach $0.49$ quickly during the course of energy minimization. A large scale analysis of protein structures show that the ratio for real and well-designed proteins is universal and equal to $0.491 \pm 0.005$. The fractional composition of hydrophobic and hydrophilic residues does not affect the ratio substantially. The ratio also holds for intrinsically disordered proteins, while it ceases to be universal for polypeptides with bad folding properties.
\end{abstract}


\preprint{Paper published in EPL --- Europhys. Lett. 96, 68005 (2011).}

\maketitle

\section{Introduction}

In recent decades, physical methods have been widely used to study properties and structures of biopolymers \cite{glaser2001,huang2005,waigh2007}, including DNA \cite{peng1992,10EPL-DNA}, RNA \cite{07pre-RNA}, and protein \cite{hyman1995,okamoto1998,dokholyan1998,11EPL-protein}. Proteins assume specified conformations from their chemical compositions or sequences to develop biological activity and functional specificity. The corresponding three-dimensional (3D) structures are a consequence of inter- and intra-molecular interactions, in which energy minimization is the principle governing the folding tendency. In spite of various components involved in the interactions, there has been attempts to derive simple geometric factors from a variety of conformations, which can be either considered as a factor for structure validity or used as an effective constraint in folding simulation.

Geometric properties of protein molecules have been studied for
more than three decades
\cite{richards1974,chothia1975,chothia1984}.
Among others, the Ramachandran plot \cite{ramachandran1963} is a practical
criterion widely used for improving the quality of NMR or crystallographic
protein structures. In a polypeptide, the main chain N-C$_{\alpha}$ and C$_{\alpha}$-C
bonds are relatively free to rotate, and can be respectively
represented by two torsion angles. These angles can only appear in
certain combinations due to steric hindrances, which define allowed regions
of the torsion angles for secondary structures in the plot.

Furthermore, it has been found that the mean volume of an amino
acid in the interior of proteins is {\bf very close to} that of
the amino acid in crystals \cite{chothia1975,chothia1984}. With
the help of the Delaunary triangulation method, Liang and Dill
\cite{liang2001} have reported that the protein packing is
heterogeneous, and in terms of packing density, protein molecules
may be either well-packed or loosely packed. Zhang {\it et al.}
\cite{zhang2003} showed that the packing density of single domain
proteins decreases with chain length, which shares a generic
feature of random polymers satisfying loose constraint in
compactness.

Beside the Ramachandran plot and the packing density which are conclusions based
on observations, there has been theoretical models introduced to simulate properties
 of protein geometric structures. For example, Banavar and Maritan \cite{banavar2003}
  have introduced the effective backbone tube model to analyze the secondary
  structures of proteins under the constraint of minimum energy and showed that
  the tube has an effective radius of $2.7$\AA.

When a polypeptide folds, the hydrophobic effects
cause nonpolar side chains to cluster together in the protein
interior or interface, whereas polar side chains tend to maximize
the contacts with outer solvent molecules. The stability of
the system is partially due to the burial of the nonpolar
residues, and can be measured by the loss of the solvent
accessible surface area \cite{lee1971,richard1977,connolly1983,stephanie2006}. An atom or group of atoms
is defined as accessible if a solvent molecule of specified size,
generally water, can be brought into van der Waals contact. The
solvent accessible surface is then simply defined as the surface
traced out by the center of a probe sphere, which represents the
solvent molecule, as it rolls over the van der Waals surface of
the protein \cite{connolly1983,05jcc}. Hence, volume and surface area
are suitable parameters to characterize the geometrical conformation of protein.

\section{Methods} \label{methods}
\subsection{Folding simulation}
We used the SMMP package \cite{eisenmenger2001,eisenmenger2006} for protein folding simulation and simulated annealing, as well as canonical Monte Carlo method, to generate folded structures. Starting with a polypeptide in a solvent, the SMMP searches the lowest energy conformation by utilizing the energy function
\begin{equation}
E_{tot} = E_{LJ} + E_{el} + E_{hb} + E_{tors}, \label{etotal}
\end{equation}
where
\begin{eqnarray}
E_{LJ} & = & \sum _{j>i} \left( \frac{A_{ij}}{r^{12}_{ij}} - \frac{B_{ij}}{r^{6}_{ij}}\right), \\ \label{elj}
E_{el} & = & 332 \sum _{j>i} \frac{q_i q_j}{\varepsilon r_{ij}}, \\ \label{eel}
E_{hb} & = & \sum _{j>i} \left( \frac{C_{ij}}{r^{12}_{ij}} - \frac{D_{ij}}{r^{10}_{ij}}\right), \\ \label{ehb}
E_{tors} & = & \sum _n U_n \left[ 1\pm cos (k_n \phi _n ) \right]. \label{etors}
\end{eqnarray}
Here $r_{ij}$ is the distance in \AA \ between atoms $i$ and $j$. $A_{ij}$,
$B_{ij}$, $C_{ij}$, and $D_{ij}$ are parameters of the empirical
potentials. $q_i$ and $q_j$ are the partial charges on the atoms $i$ and $j$ ,
respectively, $\varepsilon =2$ is the dielectric constant of the protein interior
 space. The factor $332$ in Eq.(\ref{elj}) is used to
express the energy in kcal/mol. $U_n$ is the energetic torsion barrier of rotation
 about the bond $n$ and $k_n$ is the
multiplicity of the torsion angle $\phi _n$ \cite{banavar2003}.
The input file for SMMP is a sequence of amino acids and the
output file is in the Protein Data Bank (PDB) format
\cite{berman2000}. The protein-solvent interactions were
implemented with the implicit water solvation by selecting type 1
solvent in the SMMP \textit{main.f} program. All parameters needed
for the simulation have been self-contained in the SMMP package.

\subsection{Calculation of volume and surface area}

To compute the volume $V$ and surface area $A$ of the polypeptide
in the course of folding simulation, we used the ARVO package
\cite{05cpc} developed based on analytic equations \cite{05jcc}.
ARVO can calculate $V$ and $A$ of a system of $N$ atoms, which can
overlap in any way. The main idea of the algorithms of ARVO is
converting computation of volume and surface area of overlapping
spheres as surface integrals of the second kind over closed
regions. Using stereographic projection, one can transform the
surface integrals to a sum of double integrals which are then
reduced to curve integrals \cite{05jcc,05cpc}.
It has been shown that the Van der Waals
surface areas \cite{labanowski1996} computed by the GETAREA module
in FANTOM package \cite{ven1993,fraczkiewicz1998} and ARVO module
are consistent \cite{05cpc}. Comparing with programs implementing
different algorithms and approximations to describe geometrical
properties of atomic groups, the differences among the computed
surface area by VOLBL \cite{liang1998}, GEPOL
\cite{silla1990,silla1991} and ARVO \cite{05cpc} are less than
1\%, and the differences among the computed volumes are about 2\%
(see Refs.\cite{05cpc} and \cite{busa2009} for detailed
discussions). On the basis of analytical method, the accuracy of
the computation of volume and surface area of protein molecules by
using ARVO is superior to numerical integration which always
contains numerical errors \cite{busa2009}.

The input file for ARVO contains the coordinates $(x_i,y_i,z_i)$
of the center and radius $r_i$ of all $N$ atoms in the system,
where $1 \le i \le N$. The atoms can overlap in any way. To
calculate van der Waals surface area $A$ and volume $V$ of a PDB
protein structure, we used the coordinates of carbon (C), nitrogen
(N), oxygen (O), and sulfur (S) of the PDB data and van der Waals
radii of C, N, O, and S as input data. According to the
conventional parameter settings in protein folding simulations
\cite{hansmann1993,eisenmenger2001,hayryan2001,lin2003,eisenmenger2006,ghulghazaryan2007},
N atom has (van der Waals) radius $1.55$\AA, C atom has radius
$1.55$\AA, S atom has radius $2.00$\AA, and O atom has radius
$1.40$\AA. The relatively smaller radius of hydrogen (H) atom is
neglected; a water (solvent) molecule is represented by an O atom
with radius $1.40$\AA.  The radii of these atoms at the atomic
level are determined by the densities of the electron cloud, and
they are self-consistent with other physical quantities used in
the SMMP simulation \cite{eisenmenger2001,eisenmenger2006}.
Further, to calculate
solvent accessible surface area $A_s$ and related volume $V_s$ of
the protein structure, we added radius of the solvent $1.40$\AA~ to van
der Waals radii of C, N, O, and S, {\it i.e.} the effective radii of C, N, O, and S are 2.95\AA, 2.95\AA, 2.80\AA, and 3.40\AA~\cite{05jcc,05cpc},
respectively. The average atomic radius $\langle r \rangle$ and average effective radius $\langle r_s \rangle$ of folded structures are calculated using these radii.

\section{Results and Discussions} \label{results}

\subsection{$V/A\langle r \rangle$ ratio for the van der Waals volume $V$ and surface area $A$ and average atomic radius $\langle r \rangle$}

The ratio $R=V/A\langle r \rangle$ and the
total energy are computed in the time course of simulation. In all cases of our simulation, the final energy using canonical Monte Carlo method is lower than using simulated annealing. The results of $10$ small proteins (with $183 \leq N \leq 548$) (Table \ref{table}) reveal that $R$ approaches to $\approx 0.49$ as the energy decreases, while the resultant structures are not
necessary close to native structures. It turns out that the
energy minimization criterion is likely connected with the
geometric conformation defined by the ratio $R\simeq
0.49$.

\begin{table}[tbp]
\caption{The $V/A\langle r \rangle$ ratio for $10$ typical proteins structures. The $R^{\prime}$ is the $V/A\langle r \rangle$ ratio of the structure with a randomly chosen configuration by the SMMP package
\cite{eisenmenger2001,eisenmenger2006}, $R^{\prime\prime}$ is for
final structure after performing the folding simulation, subscript ``$a$'' stands for simulated annealing and ``$c$'' for canonical Monte Carlo, and $R$ is for the structure from PDB.}
\label{table}%
\begin{tabular}{lccccc}
\hline \hline %
PDB & $N$ & $R^{\prime}_{(a)}$ & $R^{\prime\prime}_{(a)}$ & $R^{\prime\prime}_{(c)}$ & $R$ \\ \hline
1HP9 & 183 & $0.5694$ & $0.4912$ & $0.4875$ & $0.5047$\\
1KDL & 193 & $0.5364$ & $0.4867$ & $0.4863$ & $0.4998$\\
1GCN & 246 & $0.5480$ & $0.4917$ & $0.4875$ & $0.4946$\\
1VII & 295 & $0.5560$ & $0.4876$ & $0.4878$ & $0.5058$\\
2PLH & 330 & $0.6280$ & $0.4864$ & $0.4857$ & $0.4954$\\
2OVO & 418 & $0.5549$ & $0.4981$ & $0.4891$ & $0.4933$\\
1PGB & 436 & $0.5385$ & $0.4866$ & $0.4878$ & $0.4891$\\
1HPT & 440 & $0.5726$ & $0.4879$ & $0.4871$ & $0.4928$\\
1UOY & 452 & $0.5414$ & $0.4896$ & $0.4872$ & $0.4945$\\
1UTG & 548 & $0.5785$ & $0.4780$ & $0.4858$ & $0.4896$\\
\hline \hline %
\end{tabular}
\end{table}

\begin{figure}[tbp]
\includegraphics[width=0.466\textwidth]{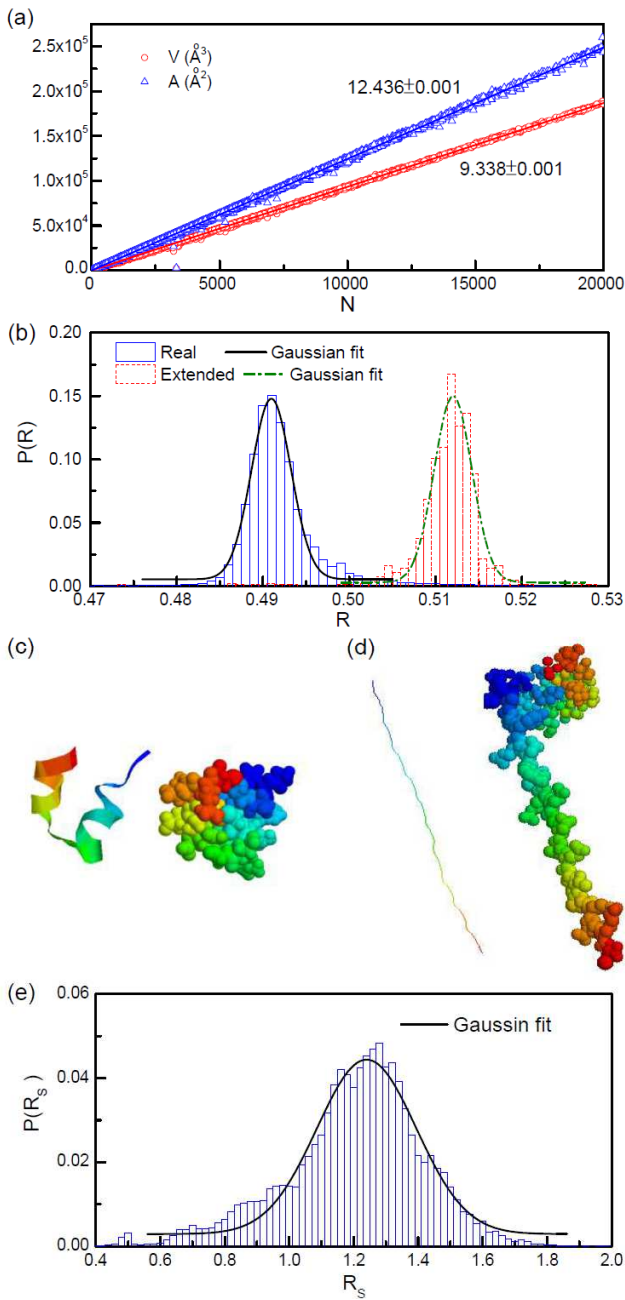} %
\caption{(a) Dependence of $V$ and $A$ on $N$ for
$28664$ PDB protein structures. The numbers indicate the
slopes. (b) The distribution $P(R)$ (blue, solid line) for the
structures shown in (a). The Gaussian fit with the maximum located
at $R \simeq 0.491$, with $y=y_0+\frac{S}{w\sqrt{\frac
{\pi}2}}exp\left[-\frac {2(x-x_c)^2}{w^2}\right] $,  $y_0 = 0.0053, x_c = 0.4910, w = 0.0046$, and $S =
0.0008$. The estimation of the fitting is adjusted
$\Re ^2 = 0.984$. The histogram of $R$ for 932 artificial extended
structured is shown in red (dotted line). The green (dash-dotted)
line refers to the Gaussian fit with maximum at $R \approx
0.5120$. (c) A typical compact PDB structure (PDB code: 1VII). (d) An extended
structure obtained by Swiss-pdb viewer 3.7 (SP5). (e) The
distribution of $R_s$ for 28236 proteins using the probe sphere
with radius of 1.4\AA. The maximum of the Gaussian fit is located
at $R_s \approx 1.2402$.}
 \label{fig1}
\end{figure}

To confirm that the ratio $R \simeq 0.49$ is relevant, we
have tested $743$ PDB structure data from
the Protein Culling Server \cite{wang2003}, in which only X-ray
data with high resolution have been selected.
The ratio is found to be $R=0.491\pm 0.005$.
To determine a reasonable tolerance for the ratio, we have also tested a
larger database from the Protein Data Bank. Totally $31059$ PDB entries deposited at the Protein Data Bank in June 2005 have been downloaded for the test. After
excluding non-proteins, such as DNA and RNA, and problematic structures in which
only $\alpha$ carbons are included, there are finally $28664$ protein structures
involved in statistics. In our analysis, both X-ray and NMR data have been
used. For NMR data consisting of more than one model, we selected
the first model which is considered as the most accurate one or is
an average of the models. We plotted the dependence of van der Waals surface area $A$ and volume $V$ on the total number of (C, N, O, and S) atoms $N$ in Fig. \ref{fig1}(a) which shows that $A$ and $V$ increase linearly with $N$. The
linear correlation between the volume $V$ and the number of atoms
$N$ or area $A$ has been found by Lorenz \textit{et al.}
\cite{lorenz1993} by using the Monte Carlo studies with the model
of clusters of random uncorrelated spheres. Similar results of the
linear relations have also been discussed by Liang and and Dill
\cite{liang2001} with $636$ protein structures. The result in Fig. \ref{fig1}(a)
provides a more solid demonstration from the basis of a larger database.
Furthermore, we plotted the distribution of
$R=V/A\langle r \rangle$ as histograms in blue (solid line) in Fig. \ref{fig1}(b), which locates in a very narrow interval centered at $R = 0.4910$.

$R \approx 0.491$ implies that one cannot imagine a protein as
a chain of small spheres because in this case we would have $R_c=
\left( 4 \pi \langle r \rangle^3/3\right)/4\pi \langle r \rangle^2\langle r \rangle = 1/3 \approx 0.333$. However, the result $R \approx 0.491$
might be understood qualitatively by considering that a protein
consists of tubes of radius $\langle r \rangle$. There is a tube
to represent the backbone of the protein; there are also some
tubes to represent side chains of the protein. The total length of
tubes is $l \sim N$. Using $V \approx \pi \langle r \rangle
^2l$ and $A \approx 2\pi \langle r \rangle l$ we obtained $V/A\langle r \rangle = 1/2 \approx 0.5$ which is consistent with
our numerical result. The linear dependence of $V$ and $A$ on $l
\sim N$ is supported by Fig. \ref{fig1}(a).
It is worth noting the linear correlation is independent of the settings
of the radii of atoms. If other radii are used, the linear
relation remains but the ratio is different. The ratio derived from the
average of an ensemble of $715$ PDB protein chains (selected by
Protein Sequence Culling Server \cite{wang2003}), using the
Richard's parameters \cite{lee1971}, is $V/A\langle r \rangle=0.5589\pm 0.0114$. Similarly, using the
Protori radii \cite{tsai1999}, the result is $V/A\langle r \rangle=0.5288\pm 0.0113$. The relation $V/A\langle r \rangle \approx
1/2$ approximately holds in the two cases.

\begin{figure}[tbp]
\includegraphics[width=0.46\textwidth]{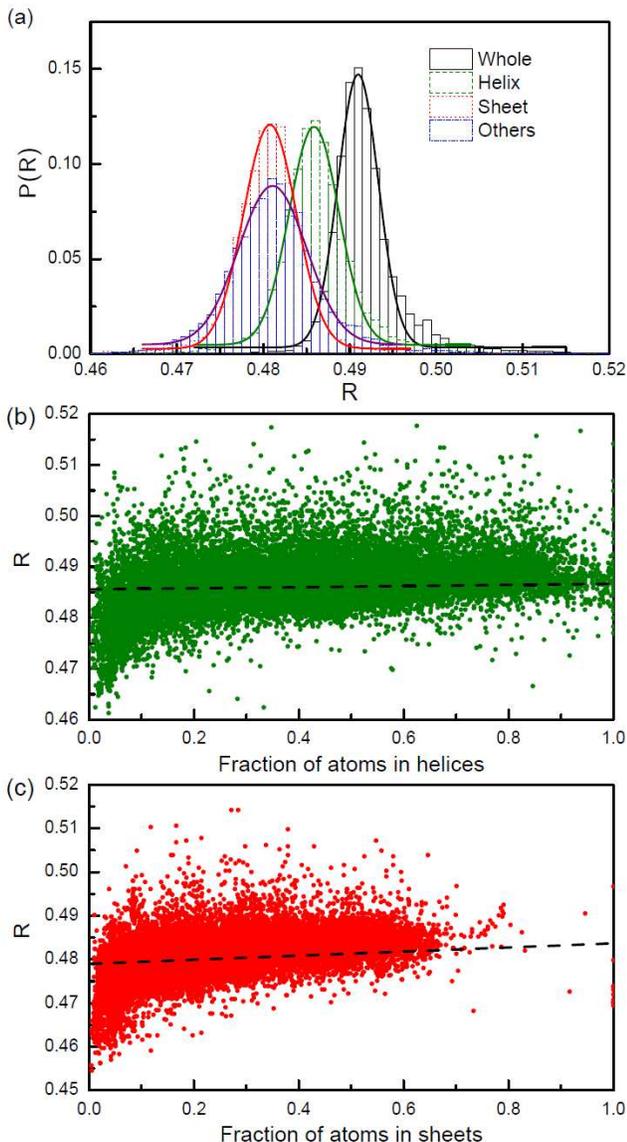} %
\caption{(a) Probability density function of $R$
statistics for whole protein molecules (28664 samples, Gaussian
distribution centered at $R=0.4910$), helix structures (extracted
from 26040 samples, $R=0.4859$), sheet structures (24537 samples,
$R=0.4808$) and other structures (25513 samples, $R=0.4811$). (b) $R$ as a function of the fraction of atoms in helix structures. The slope of the linear fit (black dashed line) is $0.0001$, and the correlation level is $0.005$. (c) $R$ as a function of the fraction of atoms in sheet structures. The slope of the linear fit
(black dashed line) is $0.0002$, and the correlation level is
$0.005$.} \label{fig2}
\end{figure}

To clarify the relation between the ratio $R \simeq
0.491$ and the compactness of native structures, we computed $R$
for artificial extended structures of protein molecules. The extended
structures are obtained by setting all torsion angles of existing 3D structures
from PDB equal to $180^\circ$, using the Swiss-pdb viewer 3.7 (SP5)
(http://www.expasy.org/spdbv/). A typical compact PDB structure and extended
structure are shown in Fig. \ref{fig1}(c) and Fig. \ref{fig1}(d),
respectively; the latter is similar to those obtained from
mechanical unfolding of proteins studied in Refs.\cite{li2006,li2007}.
The histograms in red (dotted line) in Fig. \ref{fig1}(b) show the
distribution of $R$ for 932 artificial structures. Its maximum
locates at $R \approx 0.5120$ which is higher compared to real
protein structures. This interesting result confirms that the
value $R \approx 0.491$ comes from the requirement for the
formation of compact native conformations as a result of energy minimization.

Further, direct comparisons of volumes and surface areas for real
and extended structures show that from an extended structure to a
real structure, there is a small change (increase or reduction) in
volume while there is usually a large increase in surface such
that the ratio changes from $0.512$ to $0.491$. All of these
indicate that the larger ratio of extended structure is
attributed to nonphysical geometrical properties, such as loosely
connections of monomers, and unbalance of electrostatic interactions among
monomers, and interactions between monomers and water molecules.
It should be noted that both real and artificial structures
satisfy the requirements imposed by the Ramachandran plot, but
only the former has protein-like properties. Thus, $R \approx
0.491$ can serve as an useful factor for selecting
three-dimensional protein-like structures.
In addition, we have also found that the beta structures extracted
from PDB protein structures have smaller $R=0.4808$ in
comparison with the helixes ($R=0.4859$), as shown in Fig.
\ref{fig2}(a). Whereas the ratios for individual secondary
structures are different, a protein molecule as whole is a
self-organized geometric unit, which blends various secondary
structures to form a properly folded 3D structure. In contrast to
secondary structures, tertiary and quaternary structures then have
universal property of $R \simeq 0.491$ regardless of their
details.
Defining the relative beta/helix content of a protein as a number of amino
acids belonging to beta strands/helix structures divided by its total
number of residues, we found that, as shown in Figs.
\ref{fig2}(b) and \ref{fig2}(c), there is no correlation
between $R$ and beta- as well as helix-content as the correlation
level for the linear fits is very low (0.005) for both cases.

\subsection{$V_s/A_s\langle r_s \rangle$ ratio for solvent accessible volume $V_s$ and surface area $A_s$ and average effective radius $\langle r_s \rangle$}
In order to compare the distributions of $R$ (with zero radius of
solvent) and $R_s \equiv V_s/A_s\langle r_s \rangle$ (with radius of solvent 1.4\AA), we calculated the distributions of $R$ and $R_s$ for $28236$
protein structures from PDB. The histogram of $R$ is
not shown because it is similar to the larger set of 28664 protein
structures (Fig. \ref{fig1}(b)). The distribution of $R_s$
(Fig. \ref{fig1}(e)) has a maximum at $R_s \approx 1.2402
$.

\subsection{$V/A\langle r \rangle$ ratio and hydrophobicity of amino acids}

\begin{figure}[tbp]
\includegraphics[width=0.466\textwidth]{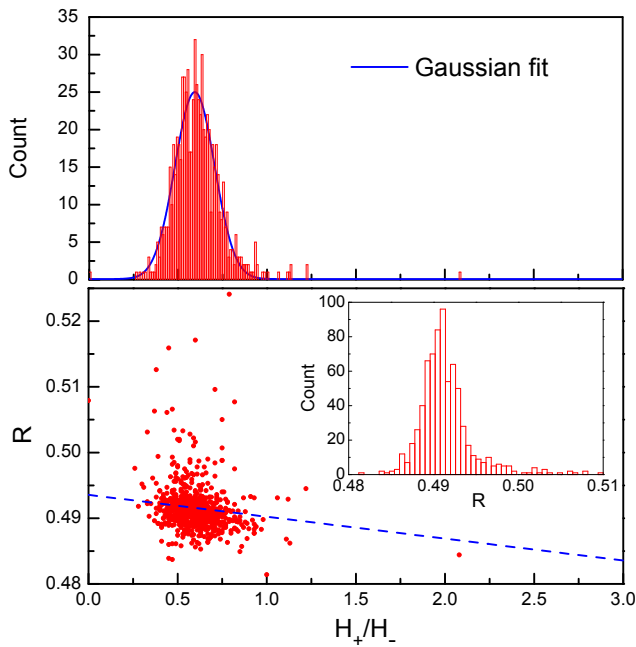} %
\caption{Upper panel shows the distribution of the
ratio of hydrophilic ($H_{+}$) and hydrophobic ($H_{-}$) amino
acids in a molecule for $723$ protein structures (from Protein
Sequence Culling Server \cite{wang2003}), based
on the Kyte-Doolittle scale \cite{kyte1982}. The Gaussian fit is
centered at $0.594$. Lower panel shows $R$ as a function of
$H_+/H_-$. The slope of the linear fit (blue dashed line) is
$-0.0033$, and the correlation level is $0.2$. The inset shows the
distribution of $R$ for the $723$ protein structures.}
\label{fig3}
\end{figure}

Consider a polypeptide chain consisting of a sequence of amino
acids with different hydrophobicities. The hydrophobic
condensation drives the polypeptide chain toward a conformation
with lower free energy. This is achieved by burying hydrophobic
contents into interior and reside polar monomers on the surface
contacting with water. This process involves not only the
regulation of the connections between monomers but also
compensations of volume and surface area.
According to the statistics shown in Fig. \ref{fig3} for $723$
protein structures (from Protein Sequence Culling Server
\cite{wang2003}), the ratio of hydrophilic and hydrophobic amino
acids ($H_+/H_-$) of proteins in the Kyte-Doolittle scale
\cite{kyte1982} is generally in a narrow range with respect to
variable range of $H_+/H_-$, suggesting that the
universality of $R$ is probably a consequence of compositions of
hydrophilic and hydrophobic amino acids in a protein.
The linear fit for $R$ as a function of $H_+/H_-$ (lower part of
Fig. \ref{fig3}) gives the correlation level of $0.2$. Since this level is
notably lower than 0.5 there is no correlation between these two
quantities. This is not unexpected because $R$ varies in a very
narrow interval. For this reason, one can show that $R$ does
not correlate with individual values of $H_+$ and $H_-$. Furthermore, folding simulations of ten polypeptides with fixed $H_+/H_-$ and randomized sequences show that the averages of the ratios are $\langle R^{\prime}\rangle=0.548\pm0.013$ for initial structure and $\langle R^{\prime \prime}\rangle=0.490\pm0.001$ for final structure after energy minimization. This implies that the ratio is not only the property of disordered protein, but is also that of random copolymers.

\subsection{$V/A\langle r \rangle$ ratio for intrinsically disordered proteins}

\begin{figure}[tbp]
\includegraphics[width=0.456\textwidth]{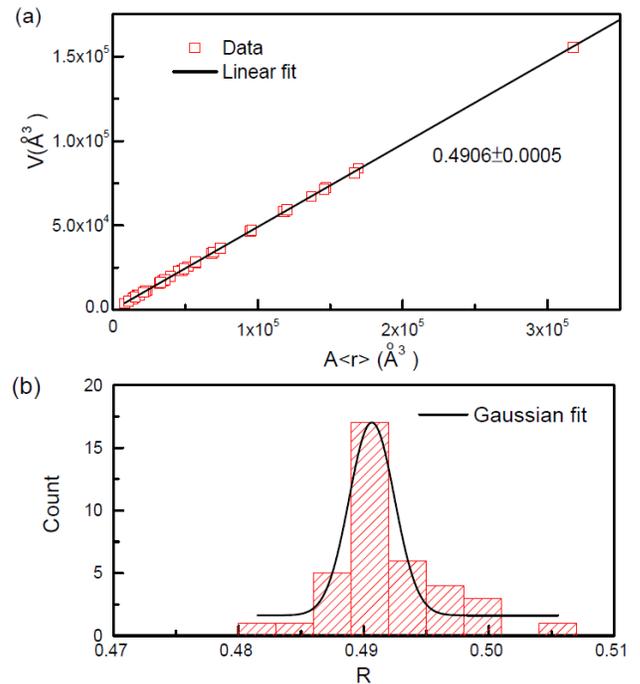} %
\caption{(a) Volume $V$ as a function
of surface area $A$ for $38$ disordered proteins
\cite{meszaros2007}. (b) The histogram of the structures shown in
(a). The Gaussian fit with the maximum located at $R \simeq 0.4906
$.} \label{fig4}
\end{figure}

It is also of interest to study the ratio for the intrinsically
disordered proteins which usually lead to misfolding
\cite{meszaros2007,mittag2007}. We have calculated $R$ for $38$
protein structures \cite{meszaros2007} and found that the ratio is
$0.4906 \pm 0.005$(see Fig. \ref{fig4}), which is within the
tolerance determined by the ensemble of 28664 PDB protein
structures. Thus, the ratio holds once a polypeptide fold to a compact structure no matter of its species.

\section{Summary and Conclusions} \label{summary}

In summary, we have found a universal ratio of the van der Waals
volume to the surface area and average atomic radius of folded structures $R = 0.491 \pm 0.005$ for native
protein structures,
including intrinsically disordered proteins.
We have studied the connection between the energy minimization and
geometric conformation by monitoring the ratio $R$ during folding
simulations using the SMMP package
\cite{eisenmenger2001,eisenmenger2006}. Our results reveal that $R
\simeq 0.491$ should be somewhat related to the energy global
minimum of protein molecules. This result can be imposed as a rule
in searching for native conformations in folding simulations using
protein sequences. $R \approx 0.491$ can also serve as a
necessary condition for checking the validity of PDB data and
designing protein-like sequences.

It is well known that hydrophobic residues are buried in the core
of proteins and the van der Waals volume should be, therefore,
proportional to the number of such residues. The van der Waals
area should linearly depend on the number of hydrophilic residues,
which have tendency to reside on the protein surface. Thus, the
universality of $R$ is probably a consequence of the fact that the
ratio of the hydrophobic and hydrophilic amino acids of proteins
is roughly a constant.

Here we should emphasize that $R \simeq 0.491$ does not
correspond to a unique conformation, but it confines molecular
conformations in a folding simulation from vast possibilities to a
smaller space. It excludes improperly folded structures which are
characterizable by such geometrical properties and is beneficial
for the reduction of simulation time.
One possible implementation of this property shall be in the
calculation of surface energy associated with the solvent access
area. A preliminary test of the $V/A\langle r \rangle$, working as a filter, can be performed  before next update step in simulations. Other independent factors can work together to define the conformation to have a native-like structure.

\begin{acknowledgments}
This work was supported by Taiwan Grants NSC 96-2112-M-008-021-MY3, 96-2911-M-001-003-MY3, 97-2627-B-008-004, 98-2627-B-008-004, 99-2627-B-008-002, 100-2112-M-008-003-MY3, Poland-Taiwan NSC Grant 100-2911-I-001-507, NCTS (North), AS-95-TP-A07, and the Polish Grant No. 202-204-234.
\end{acknowledgments}

\end{document}